\documentclass[12pt]{article}

\catcode`\@=11

\global\arraycolsep=2pt
\usepackage{amsbsy,amssymb,latexsym,amsfonts,amsmath}
\usepackage{graphicx,color}

\newcommand\red[1]{{\textcolor{red}{#1}}}

\begin{document}
\begin{flushright}
\parbox{4.2cm}
{RUP-17-13}
\end{flushright}

\vspace*{0.7cm}

\begin{center}
{ \Large Very Special Conformal Field Theories (VSCFT) and their holographic duals }
\vspace*{1.5cm}\\
{Yu Nakayama}
\end{center}
\vspace*{1.0cm}
\begin{center}

Department of Physics, Rikkyo University, Toshima, Tokyo 171-8501, Japan

\vspace{3.8cm}
\end{center}

\begin{abstract}
Cohen and Glashow introduced the notion of very special relativity as viable space-time symmetry of elementary particle physics. As a natural generalization of their idea, we study the subgroup of the conformal group, dubbed very special conformal symmetry, which is an extension of the very special relativity. We {classify all of them} and construct field theory examples as well as holographic realization of the very special conformal field theories.
\end{abstract}

\thispagestyle{empty} 

\setcounter{page}{0}

\newpage




\section{Introduction}
To a very good approximation, the Poincar\'e group (i.e. space-time translation and Lorentz transformation) is symmetry of particle physics. However, it may not be the fundamental symmetry of the nature. The gravity certainly affects the space-time symmetry. Even without gravity, the Lorentz symmetry may be just accidental low energy effective symmetry, or alternatively it may be violated at a very large distance. In both cases, the nature may not preserve the Lorentz symmetry but it is violated with a tiny amount that has not been observed yet.

Sometime ago Cohen and Glashow introduced the notion of very special relativity as viable space-time symmetry of elementary particle physics \cite{Cohen:2006ky}\cite{Cohen:2006ir}. They proposed that a certain subgroup of the Poincar\'e symmetry may be the fundamental space-time symmetry. The group spanned by a subgroup of the Poincar\'e generator $P_+$, $P_-$, $P_i$, $J_{+i}$ is dubbed very special relativity. A crucial observation there is that if we added $CP$ symmetry, it would be enhanced to the Poincar\'e group because it exchanges $+$ and $-$ so that we need to add $J_{-i}$ and then the commutation relations imply the existence of $J_{ij}$. 
Since the $CP$ violation of the standard model is somehow small, the violation of the Lorentz symmetry down to the very special relativity might be explained to be naturally small as we observe.

Before the advent of the special relativity by Einstein, the Poincar\'e group had been known to be symmetry of the Maxwell theory of electrodynamics. Actually, the Maxwell theory in the vacuum (without source) has a larger symmetry known as the conformal symmetry (but only in $1+3$ dimensions \cite{ElShowk:2011gz}). In this article, we ask the question what is a subgroup of the conformal group that is consistent with the very special relativity. Once we determine it, it is important to ask the further question what are the actual field theory realizations. In this paper, we will give examples of what we call very special conformal field theories which are field theories with no larger space-time symmetry than the very special conformal symmetry.

Conformal field theories are natural candidates of renormalization group fixed points, and they play significant roles in our understanding of critical phenomena. They may also play important roles in particle physics. This is because most of the scale invariant field theories are actually conformal (see e.g. \cite{Nakayama:2013is} for a review), and at the high or low energy limit, the physics will be governed by renormalization group fixed points without any intrinsic scale.
We then expect that the very special conformal field theories should play a crucial role in our understanding of the very special relativity. This leads us to the next immediate question if the scale invariant very special relativity naturally leads to very special conformal field theories. 
For this purpose, we will study the structure of the energy-momentum tensor by focusing on the possibility of the improvement. Another question is if the very special conformal field theories must have secret symmetry enhancement to the full Poincar\'e conformal symmetry.  We will address these issues by offering examples as well as general argument.

In recent years, it has become more and more common to study strong dynamics of quantum field theories by using holography, even without the Poincar\'e invariance. In this paper, we also construct a holographic dual description of very special conformal field theories. In particular, in some versions of the very special conformal symmetry it is possible to construct a holographic model not only in the effective gravity, but also in the full string theory background, so we naturally believe that the very special conformal field theories are not theories in the swampland but are consistent by themselves. In other versions of the very special conformal symmetry, it is an open question if we can construct any holographic dual description.

The organization of the paper is as follows. In section 2, we define very special conformal symmetry and study the properties of the energy-momentum tensor. In section 3 we propose field theoretic descriptions of very special conformal field theories, and in section 4 we show some examples. In section 5, we study holographic dual descriptions of very special conformal field theories. In section 6, we conclude with discussions.

\section{Very special conformal symmetry}
Let us first fix our convention. We use the light-cone coordinate: $x^+ = \frac{1}{\sqrt{2}}(t+x)$, $x^-=\frac{1}{\sqrt{2}}(t-x)$ and $x^i$ $(i=y,z)$, or $x_+ = -\frac{1}{\sqrt{2}}(t-x)$, $x_- = -\frac{1}{\sqrt{2}}(t+x)$ and $x_i=x^i$. Similarly, we define the light-cone vectors as  $ A^+ = \frac{1}{\sqrt{2}}(A^0+A^x)$, $A^-=\frac{1}{\sqrt{2}}(A^0-A^x)$ or $A_+ = -\frac{1}{\sqrt{2}}(A^0-A^x) = \frac{1}{\sqrt{2}}(A_0+A_x)$, $A_- =-\frac{1}{\sqrt{2}}(A^0+A^x)=\frac{1}{\sqrt{2}}(A_0-A_x)$. The light-cone derivatives are defined as $\partial_+ = \frac{\partial}{\partial x^+} = \frac{1}{\sqrt{2}}(\partial_t + \partial_x)$ and $\partial_- = \frac{\partial}{\partial x^{-}} = \frac{1}{\sqrt{2}}(\partial_t - \partial_x)$. We note that $A_\mu B^\mu = A^+B_+ + A^-B_- + A^i B_i = -A^+B^- - A^-B^+ + A^iB^i= -A_+B_- - A_+ B_- + A_i B_i$. 

The very special relativity is  symmetry of the space-time, which is a subgroup of the Poincar\'e group, given by $P_+$, $P_-$, $P_i$, $J_{+i} = \frac{1}{\sqrt{2}}(J_{ti}+J_{xi})$, where $P_\mu$ is the space-time translation and $J_{\mu\nu}= -J_{\nu\mu}$ is the Lorentz transformation. This is the minimal version of the very special relativity, and apart from the space-time translation, the Lorentz part of the group is $T(2)$. There are a couple of different extensions of the very special relativity. Firstly, we may supplement $J_{ij}$, and the Lorentz part becomes $E(2)$. Secondly, we may supplement $J_{+-}$ instead, and the Lorentz part becomes $HOM(2)$. Finally, we may supplement both $J_{ij}$ and $J_{+-}$ and the Lorentz part becomes $SIM(2)$.\footnote{The total very special relativity algebra has various names in the literature. The combination of $E(2)$ and $P_\mu$ is sometimes called the Bargmann algebra or massive Galilean algebra (see e.g. \cite{Andringa:2010it} and reference therein). The combination of $SIM(2)$ and $P_\mu$ is called $ISIM(2)$ algebra in \cite{Gibbons:2007iu}.}

 
The very special conformal symmetry is defined by adding two operators $\tilde{D} = D+ J_{xt}$ and $K_+ = \frac{1}{\sqrt{2}}(K_t+K_x)$, where $D$ is the generator of the Poincar\'e dilatation and $K_\mu$ is the generator of the Poincar\'e special conformal transformation. We first argue that this is the only possibility in any version of the very special relativity. Suppose we add $K_i$ to the very special relativity, then the commutator of $K_i$ and $P_-$ gives $J_{-i}$. Similarly the commutator of $K_i$ and $P_j$ gives $J_{ij}$ as well as $D$. The commutator of $J_{-i}$ and $J_{+j}$ gives $J_{+-}$, so the entire Poincar\'e symmetry is recovered. Instead, suppose we add $K_-$, then the commutator of $K_-$ and $J_{+i}$ gives $K_{i}$ and the above argument follows. Thus adding $K_+$ is the only special conformal symmetry that is consistent with the very special relativity.

We list the schematic form of the commutator $i[X,Y]$ in table 1 in the most extended case of $SIM(2)$ with the very special conformal symmetry. The other special conformal symmetry with $T(2)$, $E(2)$ and $HOM(2)$ can be obtained just as a subgroup by neglecting $J_{ij}$, $J_{+-}$ or both in each case. Since the commutator with $K_{+}$ does not give rise to them, these are all consistent subalgebras.

\begin{table}[htb]
  \begin{tabular}{|c|cccccccc|} \hline
     & $P_+$ & $P_-$ & $P_i$ & $J_{ij}$ & $J_{+i}$ & $J_{+-}$ & $K_+$ & $\tilde{D}$ \\ \hline 
$P_+$ & $0$ & $0$ & $0$ & $0$ & $0$ & $-P_+$ & $0$ & $0$ \\ 
$P_-$ & $0$ & $0$ & $0$ & $0$ & $P_i$ &$ P_-$ & $-\tilde{D}$ & $2P_-$ \\ 
$P_i$ & $0$ & $0$ & $0$ & $P_j$ & $P_+$ & $0$ &$J_{+i}$ & $P_i$ \\ 
$J_{ij}$ & $0$ & $0$ & $-P_j$ & $J_{kl}$ & $J_{+i}$ & $0$ & $0$ & $0$ \\ 
$J_{+i}$ & $0$ & $-P_i$ & $-P_+$ & $-J_{+i}$ & $0$ &$-J_{+i}$ & $0$ & $-J_{+i}$ \\ 
$J_{+-}$ & $P_+$ & $-P_-$ & $0$ & $0$ & $J_{+i}$ &$0$ & $K_+$ & $0$ \\ 
$K_+$ & $0$ & $\tilde{D}$ & $-J_{+i}$ & $0$ & $0$ &$-K_+$ & $0$ & $-2K_+$ \\ 
$\tilde{D}$ & $0$ & $-2P_-$ & $-P_i$ & $0$ & $J_{+i}$ &$0$ & $2K_+$ & $0$ \\ \hline
  \end{tabular}
\caption{The commutation relation of very special conformal generators.}
\end{table}

We should note that in the case of $T(2)$ and $E(2)$, $P_+$ is the center of the algebra. Furthermore, the very special conformal algebra with $E(2)$ is isomorphic to the Schr\"odinger algebra in $1+2$ dimensions \cite{Nishida:2007pj}.

In order to formulate field theoretic realizations of the very special conformal symmetry, it is convenient to study the properties of the energy-momentum tensor.\footnote{We work on the Noether or canonical construction of the energy-momentum tensor, but an equivalent way is to couple the theory to non-relativistic Newton-Cartan gravity and study the gravitational source  as in \cite{Jensen:2014aia}\cite{Hartong:2014pma}. This has been mostly studied in the $E(2)$ case.}
First of all, the space-time translation requires the existence of the conserved energy-momentum tensor $T_{\mu}^{\ \nu}$:
\begin{align}
\partial_+ T_{+}^{\ +} + \partial_- T_{+}^{\ -} + \partial_i T_+^{\ i} &= 0 \cr
\partial_+ T_{-}^{\ +} + \partial_- T_{-}^{\ -} + \partial_i T_-^{\ i} &= 0 \cr
\partial_+ T_{j}^{\ +} + \partial_- T_{j}^{\ -} + \partial_i T_j^{\ i} &= 0
\end{align}
so that the charges for the space-time translations $P_+ = \int dt T_+^{\ 0}$, $P_- = \int dt T_-^{\ 0}$, $P_i = \int dt T_i^{\ 0}$ are conserved.

In the very special relativity, the existence of the very special Lorentz transformation $J_{+i}$ demands the conservation of the very special Lorentz current $J_{+i}^{\mu}$:
\begin{align}
J_{+i}^+ &= -x^{-} T_i^{\ +} - x_i T_{+}^{\ +} \cr
J_{+i}^- &= -x^{-} T_{i}^{\ -} - x_i T_{+}^{\ -} \cr
J_{+i}^j &= -x^{-} T_{i}^{\ j} - x_i T_{+}^{\ j} \ , 
\end{align}
from which we require $-T_i^{\ -} - T_{+}^{\ i} = 0$. If $J_{ij}$ is conserved,  we further require $T_{i}^{\ j} = T_{j}^{\ i}$. Similarly, if $J_{+-}$ is conserved, we require $T_{+}^{\ +} = T_{-}^{\ -}$. 

With the very special conformal symmetry, we further demand the conservation of the dilatation current $\tilde{D}^\mu$
\begin{align}
\tilde{D}^+ & = (2x^{-}) T_{-}^{\ +} + x^i T_{i}^{\ +} \cr
\tilde{D}^- & = (2x^{-}) T_{-}^{\ -} + x^i T_{i}^{\ -} \cr
\tilde{D}^j &= (2x^{-}) T_{-}^{\ j} + x^i T_{i}^{\ j} 
\end{align}
with $\tilde{D} = \int dt \tilde{D}^{0}$. Furthermore the very special conformal current must be conserved:
\begin{align}
K_+^+ &= (x^-)^2 T_{-}^{\ +} + x^-x^i T_{i}^{\ +} + \frac{(x_i)^2}{2} T_{+}^{\ +} \cr
K_+^- &= (x^-)^2 T_{-}^{\ -} + x^-x^i T_{i}^{\ -} + \frac{(x_i)^2}{2} T_{+}^{\ -} \cr
K_+^j &= (x^-)^2 T_{-}^{\ j} + x^-x^i T_{i}^{\ j} + \frac{(x_i)^2}{2} T_{+}^{\ j}
\end{align}
For the conservation of $K_+ = \int d^3x K_+^0$, it is sufficient that the energy-momentum tensor is ``traceless": $2T_{-}^{\ -} + T_{i}^{\ i} = 0$ and symmetric $-T_i^{\ -} - T_{+}^{\ i} = 0$.

The traceless condition is not necessary because the energy-momentum tensor has an ambiguity in its definition. For instance, it allows the improvement of the form $\tilde{T}_{\mu}^{\ \nu} = T_{\mu}^{\ \nu} + \frac{1}{3}(\partial_\mu \partial^\nu -\delta_\mu^\nu \partial^\rho \partial_\rho) L$ with a certain local operator $L$. Thus, when $ 2T_{-}^{\ -} + T_{i}^i = (-2\partial_+ \partial_- + \partial_i^2 ) L$, one can always introduce the traceless energy-momentum tensor $\tilde{T}_{\mu}^{\ \nu}$ so that the very special conformal symmetry is realized. 


Finally, we note that while the very special conformal symmetry is the only non-trivial extension of the very special relativity that contains the ``conformal transformation", there are many other possibilities with only extra ``dilatation" symmetry with $\tilde{D}_{\lambda} = D+ \lambda J_{xt}$. The very special conformal symmetry $K_+$ is compatible only with $\lambda=1$ (or $\lambda=0$ in which case we recover the full Poincar\'e conformal symmetry).

\section{Field theory deformations to very special conformal field theories}
In this section we discuss general recipes to construct very special conformal field theories from the Poincar\'e invariant conformal field theories by adding local operators. The construction in particular applies to Lagrangian field theories because they are based on the free conformal field theories. In the next section we will show some examples.

Let us first consider easier cases with $E(2)$ and $T(2)$ symmetry, which we can construct via local non-singular deformations. 
In the $E(2)$ case, we may start with a conformal field theory by adding the ``irrelevant" deformation
\begin{align}
\delta S = \lambda \int d^3x dt J_+ \ , \label{e2}
\end{align}
where $J_+$ is a vector primary operator with Poincar\'e scaling dimension $\Delta_J = 5$. Here the ``irrelevance" is with respect to the Poincar\'e dilatation $D$ and the deformation is actually marginal with respect to $\tilde{D}$.\footnote{\red{The different notion of relevance between $D$ and $\tilde{D}$ is also emphasized in \cite{Guica:2010sw}, where they consider the similar deformations of Poincar\'e conformal field theories to obtain Schr\"odinger invariant field theories.}} 
Indeed, one may easily check that the added operator is invariant under $\tilde{D}$ and $K_+$  at the classical level, but breaks the Lorentz symmetry down to $E(2)$. One may generalize the construction with rank $n$ tensor $J_{++\cdots}$ with the Poincar\'e scaling dimension $\Delta = 4+n$.

In the $T(2)$ case, we instead consider adding the ``irrelevant deformation"
\begin{align}
\delta S = \lambda \int d^3x dt A_{+x} \ , \label{t2}
\end{align}
where $A_{+x} = -A_{x+}$ is an anti-symmetric tensor with the Poincar\'e scaling dimension $\Delta_A = 5$. At the classical level, the added operator is invariant under $\tilde{D}$ and $K_+$, but breaks the Lorentz symmetry down to $T(2)$. In particular, it is not invariant under $J_{ij}$. One may consider the higher rank tensor deformations that have more anti-symmetric $[+x]$ and $+$ indices. 

The case with $SIM(2)$ and $HOM(2)$ is much more non-trivial. In \cite{Cohen:2006ky}, it was argued that there is no local field theory constructions because of the lack of the non-trivial spurion field. They instead considered the non-local filed theory examples. We will discuss the conformal extension in section 4.4.
Here we would like to propose local but singular constructions that at least make sense classically. 
In the $SIM(2)$ case, we consider the deformation
\begin{align}
\delta S = \lambda \int d^3x dt \frac{J_+}{\tilde{J}_{+}} \ ,
\end{align}
where $J_\mu$ and $\tilde{J}_\mu$ are primary vector operators, and the Poincar\'e scaling dimensions must obey $\Delta_J - \Delta_{\tilde{J}} = 4$. Compared with \eqref{e2}, this deformation preserves the additional $J_{+-}$ symmetry.
There is an obvious generalization with higher rank tensors, which we will not dwell on.

Similarly, in the $HOM(2)$ case, we consider the deformation
\begin{align}
\delta S = \lambda \int d^3x dt \frac{A_{+x}}{\tilde{A}_{+x}} \ ,
\end{align}
where $A_{\mu\nu}$ and $\tilde{A}_{\mu\nu}$ are primary anti-symmetric tensor operators, and the Poincar\'e scaling dimensions must obey $\Delta_A - \Delta_{\tilde{A}} = 4$. 

These deformations are singular in the sense that the inverse of the operator such as $1/\tilde{J}_+$ may not be well-defined quantum mechanically. We, however, point out that in quantum mechanics, the inverse of the position operator is frequently used in conformal quantum mechanics (as well as in the Coulomb potential problem!). It is however possible that with these singular deformations, the ground state may break the very special conformal symmetry spontaneously.

It is not entirely impossible to construct $SIM(2)$ invariant very special conformal field theories by using non-local interactions as proposed in \cite{Cohen:2006ky}. As we have already mentioned, since there is no obvious general argument, we will separately show examples  in the next section.

\section{Some examples}

\subsection{An example: $E(2)$ case}
Let us consider a field theory model with a complex scalar $\phi$ having the action 
\begin{align}
S = \int d^3x dt & \left(-\partial_\mu \phi^* \partial^\mu \phi 
- i\lambda^\mu|\phi|^2(\phi^* \partial_\mu \phi - \phi \partial_\mu \phi^* )\right)\cr 
=  \int d^3x dt &\left( \partial_+\phi^* \partial_-\phi + \partial_+\phi \partial_-\phi^* - \partial_i \phi^* \partial_i \phi \right. \cr
   & \left. - i\lambda|\phi|^2(\phi^* \partial_+\phi - \phi \partial_+\phi^* )\right)  \ , \label{modele2}
\end{align}
where $\lambda^\mu = (\lambda^+,\lambda^-,\lambda^y,\lambda^z) = (\lambda ,0,0,0)$ is a light-like vector invariant under the special relativity.

The model is invariant under the space-time translation  $P_+$, $P_-$, $P_i$ and very special Lorentz transformation  that makes $\lambda^\mu \partial_\mu = \lambda \partial_+$ invariant i.e. $J_{+i}$ and $J_{ij}$. It is also invariant under the dilatation $\tilde{D}$ under which
\begin{align}
\delta_{\tilde{D}} \phi = (2x^{-}\partial_- + x^i\partial_i + \Delta)\phi \ , 
\end{align}
where $\Delta=1$. 
In addition, it is invariant under the very special conformal transformation
\begin{align}
\delta_{K_+} \phi = \left(2x^- \Delta + 2(x^-)^2 \partial_- + 2x^-x^i\partial_i + x_i^2 \partial_+ \right)\phi \ .
\end{align}

This model can be regarded as an example of the deformation \eqref{e2}. Here, we take the original theory as a theory of free massless  complex scalar $\phi$, and take the deformation $J_+ = |\phi|^2(\phi^*\partial_+\phi - \phi \partial_+\phi^*)$. $J_+$ has the Poincar\'e conformal dimension $\Delta_J = 5$, and by checking the conformal transformation, one can see that it is a primary vector operator.

While we may check the invariance of the action directly, we may also compute the  energy-momentum tensor:
\begin{align}
T_{\mu}^{\ \nu} =& -\partial_\mu\phi^* \partial^\nu \phi - \partial_\mu \phi \partial^\nu \phi^* \cr
& - \delta_{\mu}^\nu (-\partial_\rho\phi^*\partial^\rho \phi - i \lambda^\rho|\phi|^2(\phi^* \partial_\rho \phi - \phi \partial_\rho \phi^*)  \cr
&-i\lambda^\nu |\phi|^2(\phi^* \partial_\mu \phi - \phi\partial_\mu \phi^*) \ .
\end{align}
It has the desired properties that ensure the very special conformal symmetry discussed in the previous section. In particular, note that $T_i^{\ -} = -T_+^{\ i}$ and $2T_-^{\ -} + T_{i}^{\ i} = (-2\partial_+ \partial_- + \partial_i^2)|\phi|^2$ up to the use of the equations of motion so that one can improve the energy-momentum tensor to make it traceless. 

Let us note that the interaction is irrelevant in the Poincar\'e sense. If we take the infrared limit with the Poincar\'e scaling (i.e. $(x^+,x^-,x^i) \to \lambda (x^+,x^-,x^i)$ and $\lambda \to \infty$), the interaction vanishes. However, in the special conformal scaling limit (i.e. $(x^+,x^-,x^i) \to (x^+,\lambda^2 x^-,\lambda x^i)$ and $\lambda \to \infty$), it survives. In other word, the violation of the Poincar\'e invariance may be regarded as a UV effect.
To see this point more clearly, let us note that in the conventional $\lambda |\phi|^4$ interaction gives the $\phi(k^1)\phi^*(k^2) \to \phi(k^3) \phi^*(k^4)$ scattering amplitude of $\delta^4(k^1+k^2+k^3+k^4)$, while our interaction in \eqref{modele2} gives $\delta^{4}(k^1+k^2+k^3+k^4) (k_+^1  -k_+^2)$, which vanishes in the Poincar\'e IR limit $k_+ \to 0$, but it does not vanish in the very special conformal limit where $k_+$ is fixed. Beyond the tree level, we have to study the renormalizaiton group equation with respect to our dilatation $\tilde{D}$ rather than the Poincar\'e one $D$. In some situations one can argue that the deformations are marginal as studied in \cite{Guica:2010sw}. In our holographic example in section 5, the coupling constant is again marginal, so in these cases, the very special conformal invariance is preserved even quantum mechanically.

\subsection{An example: $T(2)$ case}
In order to obtain very special conformal field theories with the $T(2)$ symmetry, we need to add an anti-symmetric tensor with the Poincar\'e dimension $5$.  Since we have no such primary operator out of one complex scalar, we consider introducing an additional real scalar field $\varphi$ and consider the action
\begin{align}
S = \int d^3x dt & \left(-\partial_\mu \phi^* \partial^\mu \phi  -\frac{1}{2}\partial_\mu \varphi \partial^\mu \varphi - i\lambda^{\mu\nu}( \varphi (\partial_\mu \phi^* \partial_\nu \phi - \partial_\nu \phi^*\partial_\mu \phi)) \right. \cr
& \ +i \lambda^{\mu\nu} (\partial_\mu \varphi(\phi^*\partial_\nu \phi - \phi \partial_\nu \phi^*) - \partial_\nu \varphi (\phi^*\partial_\mu\phi - \phi\partial_\mu \phi^* ))\biggr) \ ,
\end{align}
where $\lambda^{\mu\nu}$ is an anti-symmetric tensor that has a non-zero component only in $\lambda^{+x} = -\lambda^{x+} = \lambda$. 
Actually, the two interaction terms are identical upon integration by part, but the combination is judiciously chosen so that it is a conformal primary operator. This is an example of deformations of \eqref{t2}: the introduced interaction $ \lambda A_{+x} =  - i\lambda^{\mu\nu}( \varphi (\partial_\mu \phi^* \partial_\nu \phi - \partial_\nu \phi^*\partial_\mu \phi))  + i \lambda^{\mu\nu} (\partial_\mu \varphi(\phi^*\partial_\nu \phi - \phi \partial_\nu \phi^*) - \partial_\nu \varphi (\phi^*\partial_\mu\phi - \phi\partial_\mu \phi^* ))$ has the  Poincar\'e dimension $\Delta_A=5$,  and by checking the conformal transformation, one can see that it is an anti-symmetric tensor primary operator.

Again we may check the invariance of the action under $K_+$ directly, but we can also compute the  energy-momentum tensor:
\begin{align}
T_{\nu}^{\ \mu} =& -\partial^\mu \phi^* \partial_\nu \phi - \partial_\nu\phi^* \partial^\mu \phi - \partial^\mu \varphi \partial_\nu \varphi \cr
&-i\lambda^{\mu\alpha}\varphi(\partial_\nu\phi^*\partial_\alpha \phi - \partial_\alpha\phi^* \partial_\nu\phi) - i\lambda^{\alpha\mu} \varphi (\partial_\alpha^*\phi \partial_\nu\phi - \partial_\nu\phi^* \partial_\alpha \phi) \cr
&+i\lambda^{\alpha\mu}(\partial_\alpha \varphi (\phi^*\partial_\nu \phi - \phi \partial_\nu \phi^*) - \partial_\nu \varphi(\phi^*\partial_\alpha \phi - \phi \partial_\alpha \phi^*)) \cr
& - i\lambda^{\mu\alpha}(\partial_\alpha \varphi(\phi^*\partial_\nu \phi - \phi \partial_\nu \phi^*) - \partial_\nu \varphi (\phi^*\partial_\alpha \phi - \phi \partial_\alpha \phi^*)) \ .
\end{align}
It has the desired properties that ensure the very special conformal symmetry discussed in the previous section. In particular, note that $T_i^{\ -} = -T_+^{\ i}$ and $2T_-^{\ -} + T_{i}^{\ i} = (-2\partial_+ \partial_- + \partial_i^2)|(\phi|^2 +\frac{1}{2}\varphi^2) $ up to the use of the equations of motion so that one can improve the energy-momentum tensor to make it traceless.

\subsection{Not an example: Maxwell-Chern-Simons theory}
In the introduction, we mentioned the symmetry of the free Maxwell theory. While the free Maxwell theory admits no relevant deformations  with the Poincar\'e invariance, it admits a non-trivial relevant deformation only with the very special relativity. It takes the form of the Chern-Simons interaction with the action given by
\begin{align}
S = \int d^3x dt \left(- \frac{1}{4}F_{\mu\nu}F^{\mu\nu} + \frac{1}{2} \epsilon^{\mu\nu\rho\sigma} k_\mu F_{\rho\sigma}A_\nu \right) \ ,
\end{align}
where $k_\mu = (k_+,k_-,k_y,k_z) = (0,k,0,0)$ is a light-like vector invariant under the special relativity.

The action is obviously invariant under $P_+$, $P_-$, $P_i$, $J_{+i}$ and $J_{ij}$. In addition it is invariant under the dilatation $\tilde{D}_{-1} = D-J_{xy}$ \cite{Hariton:2006zj}. Note that this dilatation is different from $\tilde{D} = D+J_{xy}$ that is compatible with the very special conformal symmetry. Thus, the theory is invariant under the dilatation, but is not invariant under the very special conformal transformation.

 To see it, let us study the energy-momentum tensor \cite{Hariton:2006zj}
\begin{align}
T_{\mu}^{\ \nu} = \frac{1}{4}\delta_{\mu}^{\nu} F_{\rho\sigma}F^{\rho\sigma} - F^{\nu\alpha}F_{\mu\alpha} - \frac{1}{2}k_\mu \epsilon^{\nu \alpha \rho\sigma} F_{\rho\sigma} A_{\alpha} \ .
\end{align}
It is a little bit non-trivial to check the conservation $\partial_\nu T_{\mu}^{\ \nu} = 0$: We need to use the equation of motion
\begin{align}
\partial_\mu F^{\mu\nu} + k_\mu\epsilon^{\mu\nu\rho\sigma} F_{\rho\sigma} = 0 
\end{align}
and the identity 
\begin{align}
\epsilon^{\mu\alpha\rho\sigma}F_{\rho\sigma} F^\nu_{\ \alpha} = \frac{\delta^{\mu\nu}}{4} \epsilon_{\alpha\beta \rho\sigma} F^{\rho\sigma} F^{\alpha\beta} . 
\end{align}

Having understood the conservation of the energy-momentum tensor, let us go back to the space-time symmetry of this theory. The point is that the energy momentum has the properties
\begin{align}
T_i^{\ -} + T_{+}^{\ i } &= 0 \cr
2 T_{+}^{\ +} + T_{i}^{ \ i} &= 0 \cr
T_i^{\ +} + T_{-}^{\ i} & =  - \frac{1}{2}k_- \epsilon^{i \alpha \rho\sigma} F_{\rho\sigma} A_{\alpha}  \neq 0 \cr
2T_{-}^{\ -} + T_{i}^{\ i } &=   - k_- \epsilon^{- \alpha \rho\sigma} F_{\rho\sigma} A_{\alpha}  \neq 0 \ .
\end{align}
The first equation guarantees that $J_{+i}$ is conserved, and the second equation guarantee that $\tilde{D}_{-1}$ is conserved. However, the fourth equation says that it is not invariant under $\tilde{D}$, nor $K_+$. In particular, one cannot improve the energy-momentum tensor. We also note that although $\tilde{D}_{-1}$ is conserved with the traceless condition $2 T_{+}^{\ +} + T_{i}^{ \ i} = 0 $, $K_-$ is not a symmetry because the third equation does not vanish.  

A different look at this model is the scaling behavior. Unlike the ones studied in our very special conformal field theories, the deformation here is relevant in the Poincar\'e scaling (i.e. $\Delta_J = 3$). The Chern-Simons interaction makes the photon massive at long distance and therefore it is an IR effect rather than the UV effect. On the other hand, we recall that to preserve the very special conformal symmetry, we needed $\Delta_J=5$ and the deformation must have been a UV effect. This is an intuitive reason why this model does not exhibit the very special conformal symmetry even though it is scale invariant.

\subsection{Another example: non-local $SIM(2)$ case}
The non-local model studied by Cohen and Glashow can be made very special conformal invariant by replacing the explicit mass term by the vacuum expectation value of a scalar field. Consider a Weyl fermion $\psi$ coupled with a real scalar $\varphi$ with the non-local action
\begin{align}
S = \int d^4x \left( - \frac{1}{2} \partial^\mu \varphi \partial_\mu \varphi + i\bar{\psi} \gamma_\mu \partial^\mu \psi + \varphi^2 \bar{\psi}\frac{n^\mu\gamma_\mu}{n^\mu \partial_\mu} \psi  \right) \ 
\end{align}
with $n^\mu =  (n^+,n^-,n^x,n^y) = (n,0,0,0)$, which is $SIM(2)$ invariant. The idea, being of phenomenological interest, is that upon giving a vacuum expectation value to $\varphi$, this describes a massive charged fermion without breaking the $U(1)$ symmetry, which is impossible with Lorentz invariance.

Introducing non-locality, however, makes the use of the very special conformal symmetry less trivial because most of the representation theory based on the state operator correspondence does not apply. Nor does the operator production expansion, which is at the heart of the bootstrap approach to conformal field theories. On the other hand, the violation of the non-locality is governed by the very special symmetry or $CP$ violation, so it may be tamed perturbatively.

\section{Holographic model}
Let us consider a holographic model of very special conformal field theories. 
In this paper, we only work with the case with $E(2)$ symmetry and leave the other cases for future works.
We begin with the five dimensional  Einstein-Proca theory with the action
\begin{align}
S = \int d^5x \sqrt{-g} \left(\frac{1}{2}R - \Lambda - \frac{1}{4}F_{MN} F^{MN} - \frac{m^2}{2}A_M A^M \right) . 
\end{align}
with $F_{MN} = \partial_M A_N -\partial_N A_M$. This theory has a solution with
the metric
\begin{align}
ds^2 &= g_{MN}dx^Mdx^N \cr
&=-\frac{2c^2(dx^-)^2}{z^4} + \frac{-2dx^+ dx^- + dx^i dx_i + dz^2}{z^2} 
\end{align}
and the Proca field
\begin{align}
A &= A_M dx^M = -  \frac{cdx^-}{z^2} 
\end{align}
provided
\begin{align}
\Lambda = -6 \ , \ \ m^2 = 8 \ . 
\end{align}
The constant $c$ is regarded as an exactly marginal parameter that corresponds to the coupling constant $\lambda^+$ in the previous field theory construction.

The configuration of the metric and the Proca field is manifestly invariant under $P_+$, $P_-$, $P_i$, $J_{+i}$ and $J_{ij}$. It is also invariant under the dilatation $\tilde{D}$ 
\begin{align}
x^- \to \lambda^2 x^- \ , \ \ x^i \to \lambda x^i \ , \ \ z \to \lambda z \ , \ \ x^+ \to x^+ 
\end{align}
as well as the very special conformal transformation $K_+$
\begin{align}
x^- \to \frac{x^-}{1+ax^{-}} \ , \ \ x^i \to \frac{x^i}{1+a x^-} \ , \cr
 z \to \frac{z}{1+a x^{-}} \ , \ \ x^+ \to x^+ - \frac{a}{2}\frac{x_i^2 + z^2}{1+ax^{-}} \ . 
\end{align}
The background is same as the holographic dual description of non-relativistic (Schr\"odinger) conformal field theories \cite{Son:2008ye}\cite{Balasubramanian:2008dm} except that we do not compactify the $x^+$ direction. 

We should note that the background does not solve the vacuum Einstein equation without the Proca field. There is a good reason for this: as we discussed in section 2, very special conformal field theories possess the energy-momentum tensor that has more non-trivial components than the one in the Poincar\'e conformal field theory. We therefore need more dynamical degrees of freedom in the bulk setup, which is naturally achieved by the Proca field. This local argument is stronger than the global one based on the symmetry alone that gives us the form of the metric but not the matter field \cite{SchaferNameki:2009xr}.

We point out that the above effective gravity construction may be uplifted to the full string theory compactification. In particular, one may embed the solution into the type IIB supergravity by using the so-called TST transformation  \cite{Adams:2008wt}\cite{Maldacena:2008wh}\cite{Herzog:2008wg}. This technique was used in the context of the holographic dual of Schr\"odinger field theories, and the holographic dual for $E(2)$ invariant very special conformal field theories can be simply constructed by decompactifying the light-cone direction.\footnote{The decompactified theory in the type IIB supergravity is dual to dipole deformations of $\mathcal{N}=4$ super Yang-Mills theory, which has been studied, for example, in \cite{Bergman:2000cw}\cite{Alishahiha:2003ru}\cite{Guica:2017mtd}.} The supergravity solution may be unstable once we compactify the light-cone direction, but our solution is stable since we do not compactify it. This seems consistent with the stability of the dual gauge theories under the deformations we studied (see e.g. \cite{Guica:2017mtd}).

\section{Discussions}
Cohen and Glashow proposed very special relativity as a small but natural deviation from the Poincar\'e invariance. Let us revisit a possible classification of such deformations from the renormalization group viewpoint. 
Within the renormalization group paradigm, the deformations can be classified by the operators in the undeformed Poincar\'e invariant field theory. Cohen and Glashow argued that such deformations must be a vector operator in the case of $E(2)$ case and an anti-symmetric tensor in the $T(2)$ case, and none exist in the other cases without violating locality. We, instead proposed local but singular deformations. It is an interesting question if such deformations make sense quantum mechanically.

For this purpose, it would be quite important to ask if $SIM(2)$ or $HOM(2)$ invariant very special conformal field theories can be defined in the abstract operator language without relying on the Lagrangian descriptions. In particular, we might suspect that $SIM(2)$ invariant very special conformal field theories must secretly possess the full Poincar\'e conformal invariance. Then the situation is much similar to the case with the  emergence of Poincar\'e conformal invariance from the mere scale symmetry. In $d=2$ and $d=4$, unitary scale invariant field theories with Poincar\'e invariance actually implies conformal invariance (under various technical assumptions) \cite{Polchinski:1987dy}\cite{Luty:2012ww}\cite{Dymarsky:2013pqa}.

Indeed, in $d=2$, the analogue of $SIM(2)$ (or $HOM(2)$ since there is no distinction) invariance means we have $P_+$, $P_-$, $K_+$, $\tilde{D}$ and $J_{+-}$. Then with the unitarity, the condition is sufficient to guarantee that we have $K_-$ although the algebra itself does not need it. This is just a corollary of the claim that scale invariance implies conformal invariance in $d=2$. On the other hand, the analogue of $T(2)$ (or $E(2)$ since there is no distinction) invariance in $d=2$ (i.e. $P_+$, $P_-$, $K_+$, $\tilde{D}$) implies warped conformal invariance \cite{Hofman:2011zj}\cite{Hofman:2014loa} and the theory does not have to be fully conformal invariant, example of which may be found in the literature \cite{Nakayama:2013fba}.

The construction of the holographic model for $SIM(2)$ or $HOM(2)$ invariant very special conformal symmetry is challenging. The fact that the spurion cannot be introduced in the field theory analysis may be of great hindrance in constructing the holographic dual, but it is important to understand why this is the case from the operator algebra. This should give us deeper understanding of the origin of holography. We suspect that we need a drastic change of the gravity sector.
\section*{Acknowledgements}
This work is in part supported by JSPS KAKENHI Grant Number 17K14301.


\end{document}